\documentclass[aps,showpacs,superscriptaddress,preprint]{revtex4}
\usepackage{amsfonts}
\usepackage{amsmath}
\usepackage{amssymb}
\usepackage{graphicx}
\begin{document}
\title{Structural, electronic, and thermodynamic properties of UN: Systematic density functional calculations}
\author{Yong Lu}
\affiliation{Department of Physics, Beijing Normal University,
100875, People's Republic of China} \affiliation{LCP, Institute of
Applied Physics and Computational Mathematics, Beijing 100088,
People's Republic of China}
\author{Bao-Tian Wang}
\affiliation{LCP, Institute of Applied Physics and Computational
Mathematics, Beijing 100088, People's Republic of
China}\affiliation{Institute of Theoretical Physics and Department
of Physics, Shanxi University, Taiyuan 030006, People's Republic of
China}
\author{Rong-Wu Li}
\affiliation{Department of Physics, Beijing Normal University,
100875, People's Republic of China}
\author{Hongliang Shi}
\affiliation{LCP, Institute of Applied Physics and Computational
Mathematics, Beijing 100088, People's Republic of China}
\author{Ping Zhang}
\thanks{Author to whom correspondence should be
addressed. E-mail: zhang\_ping@iapcm.ac.cn} \affiliation{LCP,
Institute of Applied Physics and Computational Mathematics, Beijing
100088, People's Republic of China} \affiliation{Center for Applied
Physics and Technology, Peking University, Beijing 100871, People's
Republic of China} 
\begin{abstract}
A systematic first-principle study is performed to calculate the
lattice parameters, electronic structure, and thermodynamic
properties of UN using the local-density approximation
(LDA)+\emph{U} and the generalized gradient approximation
(GGA)+\emph{U} formalisms. To properly describe the strong
correlation in the U $5f$ electrons, we optimized the \emph{U}
parameter in calculating the total energy, lattice parameters, and
bulk modulus at the nonmagnetic (NM), ferromagnetic (FM), and
antiferromagnetic (AFM) configurations. Our results show that by
choosing the Hubbard \emph{U} around 2 eV within the GGA+\emph{U}
approach, it is promising to correctly and consistently describe the
above mentioned properties of UN. The localization behavior of 5$f$
electrons is found to be stronger than that of UC and our electronic
analysis indicates that the effective charge of UN can be
represented as U$^{1.71+}$N$^{1.71-}$. As for the thermodynamic
study, the phonon dispersion illustrates the stability of UN and we
further predict the lattice vibration energy, thermal expansion, and
specific heat by utilizing the quasiharmonic approximation. Our
calculated specific heat is well consistent with experiments.
\end{abstract}
\maketitle

\section{introduction}
Uranium nitrides have been extensively studied in experiments in
connection with their potential applications in the Generation-IV
reactors \cite{Proc}. These reactors raise a number of concerns
surrounding the issue of nuclear energy. The fission reactions
depend on fast neutrons, requiring a small core with a high power
density and very efficient heat transfer. The oxide based fuels are
therefore being involved in the ongoing research and development,
however, the nitride fuels also participate in the competition to
become the alternative materials for their superior thermal physical
properties, such as high melting point, high thermal conductivity,
and high metal density \cite{Matzke1992}, as well as the good
compatibility with the coolant (Na).

On account of these obvious importances, several studies, such as
electronic structure optimization \cite{Evarestov, Evarestov2},
magnetic properties \cite{Rafaja}, point defects \cite{Kotomin}, and
elastic constants \cite{Doorn}, have already been conducted for
uranium nitride. However, conventional density functional theory
(DFT) that apply the LDA or GGA underestimates the strong on-site
Coulomb repulsion of the 5\emph{f}-electron and, consequently,
describes UN as incorrect FM conductor instead of the experimentally
observed AFM type-I structure \cite{antiferro} at the N\'{e}el
temperature T$_{N}$=53 K. Similar problems have been confirmed in
studying other electronically correlated materials within the pure
LDA/GGA scheme. In the present work, we use the LDA/GGA+\emph{U}
method developed by Dudarev \emph{et al.} \cite{Dudarev} to
effectively remedy the failures raised by LDA/GGA in describing the
strong intra-atomic Coulomb interaction. This method has been
successfully used to study the correlated problems \cite{Sun,
Wangbt, Shi} and the total LDA/GGA energy functional is of the form
\begin{align}
E_{\mathrm{{LDA(GGA)}+U}} =E_{\mathrm{{LDA(GGA)}}}\nonumber
+\frac{U-J}{2}\sum_{\sigma}[\mathrm{{Tr}\rho^{\sigma}-{Tr}(\rho^{\sigma
}\rho^{\sigma})],}%
\end{align}
where $\rho^{\sigma}$ is the density matrix of \emph{f} states with
spin $\sigma$, while \emph{U} and \emph{J} are the spherically
averaged screened Coulomb energy and the exchange energy,
respectively.

In this paper, we have systematically calculated the lattice
parameters, electronic structure, as well as the thermodynamic
properties of UN using the above mentioned LDA/GGA+\emph{U} scheme.
We have carefully discussed how these properties are affected by the
choice of \emph{U} as well as the choice of exchange-correlation
potential. After testing the validity of the ground state by
choosing \emph{U} around 2 eV within the GGA+\emph{U} approach, we
performed a series of calculations on the electronic structures,
bonding properties, and the phonon dispersion. The lattice vibration
energy, thermal expansion, and specific heat were obtained by
utilizing the quasiharmonic approximation (QHA) based on the
first-principles phonon density of state (DOS). The rest of the
paper is organized as follows. The computational details of
first-principles are briefly introduced in Sec. II. The calculation
results are presented and discussed in Sec. III. Finally, we give a
summary of this work in Sec. IV.

\section{computational methods}
The DFT total energy calculations were carried out using the Vienna
\textit{ab initio} simulations package (VASP)
\cite{G.Kresse1,G.Kresse2} with the projected-augmented-wave (PAW)
pseudopotentials \cite{PAW} and plane waves. The exchange and
correlation effects were described within LDA and GGA
\cite{LDA,GGA}. The uranium
6\emph{s}$^{2}$6\emph{p}$^{6}$6\emph{d}$^{2}$5\emph{f}$^{2}$7\emph{s}$^{2}$
and nitrogen 2\emph{s}$^{2}$2\emph{p}$^{3}$ electrons were treated
as valence electrons. The electron wave function was expanded in
plane waves up to a cutoff energy of 500 eV and all atoms were fully
relaxed until the Hellmann-Feynman forces become less than 0.001
eV/\AA. The Monkhorst-Pack \cite{Monkhorst} 9$\times$9$\times$9 mesh
(75 irreducible \emph{k} points) was used in Brillouin zone (BZ)
integration and the corresponding electronic DOS was obtained with
15$\times$15$\times$15 (120 irreducible \emph{k} points)
\emph{k}-point mesh. The strong on-site Coulomb repulsion among the
localized U-5\emph{f} electrons was described by using the formalism
developed by Dudarev \emph{et al.} \cite{Dudarev}. In this paper,
the Coulomb parameter \emph{U} was treated as one variable, while
the parameter \emph{J} was set to 0.51 eV. Since only the difference
between \emph{U} and \emph{J} is meaningful in Dudarev's approach,
therefore, we label them as one single parameter \emph{U} for
simplicity.

\begin{figure}[ptb]
\begin{center}
\includegraphics[width=0.5\linewidth]{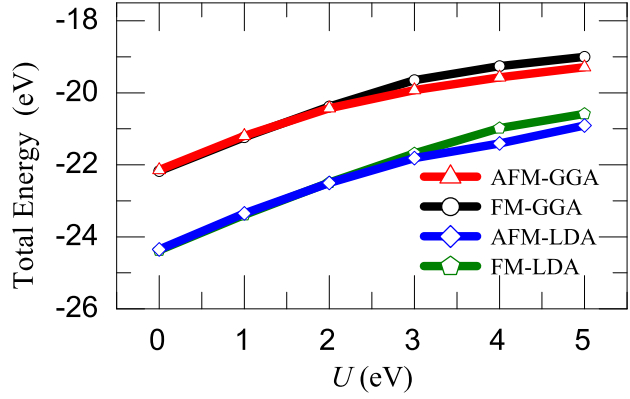}
\end{center}
\caption{(Color online) Dependence of the total energies (per formula unit) on U for FM and AFM UN.}%
\label{toten}%
\end{figure}

\section{results and discussions}
\subsection{Atomic and electronic structures of UN}
We study UN in its ground-state NaCl-type (\emph{Fm$\bar{3}$m})
structure. In the present LDA/GGA+\emph{U} approach, we have
considered the NM, FM, and AFM phases for each choice of the value
of \emph{U}, and determined the ground-state phase by a subsequent
total-energy comparison of these three phases. Compared with the FM
and AFM phases, the NM phase is not energetically favorable both in
the LDA+\emph{U} and GGA+\emph{U} formalisms. Therefore, the results
of NM are not presented in the following. The dependence of the
total energies (per formula unit) for UN in both AFM and FM
configurations on \emph{U} are shown in Fig.\ref{toten}. At
\emph{U}=0 eV, the ground state of UN is determined to be a FM
metal, which is in contrast to experiment results. By increasing the
amplitude of \emph{U}, our LDA/GGA+\emph{U} calculation correctly
predicts an AFM metal ground state and the turning value of \emph{U}
is $\sim$1.5 eV and $\sim$2 eV in GGA+\emph{U} and LDA+\emph{U}
approaches, respectively. In the discussion that follows, we,
therefore, confine our report to the AFM phase of UN.

\begin{figure}[ptb]
\begin{center}
\includegraphics[width=0.8\linewidth]{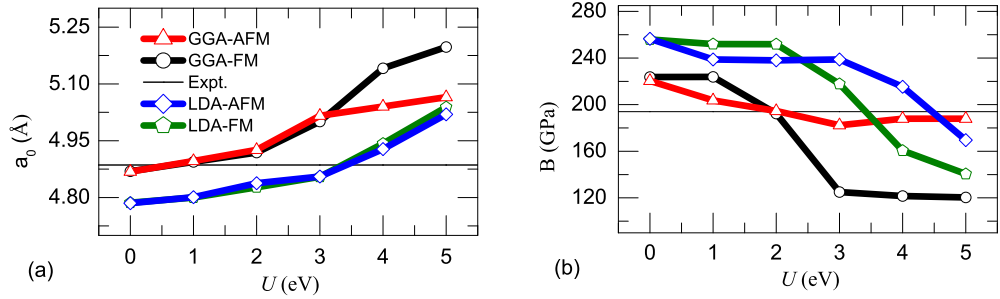}
\end{center}
\caption{(Color online) Dependence of the equilibrium lattice
parameter \emph{a$_{0}$} (a) and the bulk modulus \emph{B} (b) on
\emph{U}.}%
\label{aBU}%
\end{figure}

In this paper, the theoretical equilibrium volume \emph{V$_{0}$},
bulk modulus \emph{B} are obtained by fitting the third-order
Birch-Murnaghan equation of state (EOS) \cite{Brich}. Our calculated
lattice parameter \emph{a$_{0}$} and \emph{B} for the cubic unit
cells of UN are shown in Fig. \ref{aBU}. For the pure DFT
calculations (\emph{U}=0 eV), both the LDA and GGA methods
underestimate the lattice parameter with respect to the experimental
value. This trend is more evident for LDA approach due to its
over-binding character. After turning on the Hubbard parameter
\emph{U}, the value of \emph{a$_{0}$} gradually improves for both
LDA and GGA approaches. At around \emph{U}=1$\sim$2 eV, the
GGA+\emph{U} gives \emph{a$_{0}$}=4.896$\sim$4.926 \AA, which
consists well with the experimental data \cite{Cordfunke} of
\emph{a$_{0}$}=4.888 {\AA}. Within LDA+\emph{U}, the lattice
constant can be satisfied by turning on the Hubbard \emph{U}
parameter at around 3 eV. The dependence of bulk modulus \emph{B} on
\emph{U} is presented in Fig. \ref{aBU}(b). It is clear that the
LDA+\emph{U} results are always higher than that from GGA+\emph{U}.
This is due to above mentioned overbinding effect of the LDA
approach. With increasing the amplitude of \emph{U}, the value of
\emph{B} shows a clear declining trend for both schemes. For AFM
phase, the GGA results show that the variety of \emph{B} is small in
the range of \emph{U}=2$\sim$5 eV . At \emph{U}= 2 eV, the value of
\emph{B} equals to 194.5 GPa, which coincides well with the
experimental data (194 GPa in Ref. \cite{Matzke}, 200 GPa in Ref.
\cite{Doorn}, and 206 GPa in Ref. \cite{Olsen}). The LDA results
always hold higher \emph{B} values than experimental data till the
amplitude of \emph{U} over 4 eV.

\begin{figure}[ptb]
\begin{center}
\includegraphics[width=0.8\linewidth]{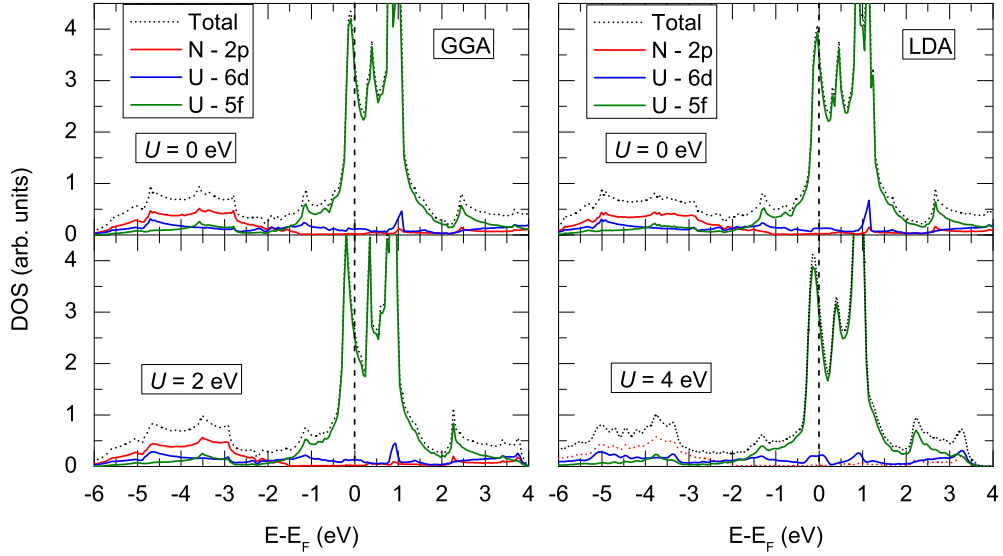}
\end{center}
\caption{(Color online) The total DOS for the UN AFM phase computed
in the GGA, GGA+U (U=2), LDA, and LDA+U (U=4) formalisms. The
projected DOSs for the U 5\emph{f}/6\emph{d} and N 2\emph{p}
orbitals are also shown. The Fermi energy level is set at zero.}%
\label{dos}%
\end{figure}

On the whole, considering the magnetic configurations, the
GGA+\emph{U} can give a satisfactory prediction of ground-state
atomic structures and bulk modulus \emph{B} by tuning \emph{U} to be
near 2 eV for UN.
Besides the above effect of LDA/GGA+\emph{U} on the atomic structure
parameters, in the following discussion we further systematically
investigate the electronic structures within the two theoretical
treatments. The total electronic DOS together with the
orbital-resolved site-projected DOS (PDOS) of UN are displayed in
Fig. \ref{dos}. Evidently, a large degree of U \emph{f} orbitals can
be observed in the valence band near the Fermi level, and the
conduction band is also strongly marked by \emph{f} orbitals. Under
the SIC-LSD calculations in Ref. \cite{Petit}, the localized and
delocalized \emph{f}-electron configurations were discussed, which
indicate that the \emph{f}$^{1}$ is the energetically favorable
configuration. However, it still remains unclear whether a
localized, delocalized, or dual localized/delocalized picture can
best account for the experimentally observed properties of UN. It is
only certain that the localization in UN is stronger than that in UC
\cite{Shi, Petit}. For UC, one part of the 5\emph{f} electrons
transfer into the interstitial zone, the other part are expected to
be confined to the \emph{j}=5/2 multiplet, and the itineracy of
5\emph{f} electrons are evident. Similar to UC, for UN, as the
increase of the Hubbard parameter \emph{U}, the split of
\emph{j}=5/2 and \emph{j}=7/2 multiplets can also be observed but
not so clear, and the itineracy of the 5\emph{f} electrons still
exists in UN. At a typical value of \emph{U}=2 eV, the conduction
band \emph{f}-electron occupancy is $\sim$2.44 electrons, compared
well with the 2.2$\pm$0.5 electrons measured by Norton \emph{et al.}
\cite{Norton} using the photoelectron-spectroscopic method. Due to
the strong overlap of the U 5\emph{f}-orbitals near the Fermi
energy, the UN phase exhibits a clear metallic behavior.

\begin{figure}[ptb]
\begin{center}
\includegraphics[width=0.3\linewidth]{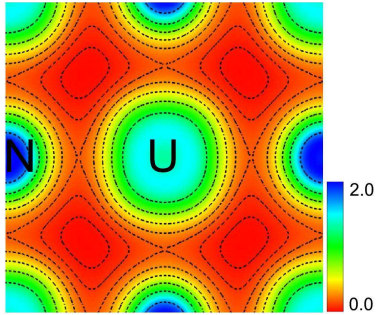}
\end{center}
\caption{(Color online) Valence charge density of UN (001) plane within GGA+\emph{U} approach at \emph{U}=2 eV.}%
\label{charge}%
\end{figure}

\begin{table}[ptb]
\caption{Bader effective atomic charges of UN. The calculated
results using LCAO (Ref. \cite{Evarestov}) and PW91 (Ref. \cite{AKotomin}) methods are also listed for comparison.}%
\begin{ruledtabular}
\begin{tabular}{cccccccccccccccc}
Methods &Bader charge&UN\\
\hline
GGA+\emph{U}&\emph{Q}$_{\texttt{U}}$&+1.71\\
&\emph{Q}$_{\texttt{N}}$&-1.71\\
LCAO &\emph{Q}$_{\texttt{U}}$&+1.58\\
&\emph{Q}$_{\texttt{N}}$&-1.58\\
PW91&\emph{Q}$_{\texttt{U}}$&+1.66\\
&\emph{Q}$_{\texttt{N}}$&-1.66\\
\end{tabular}
\label{bader}
\end{ruledtabular}
\end{table}
In order to further analyze the chemical bonding nature of UN, we
calculate the effective Bader charges \cite{Bader} in the
GGA+\emph{U} formalism with \emph{U}=2 eV. We adopt
336$\times$336$\times$336 charge density grids and the spacing
between adjacent grid points is 0.011 {\AA}. The calculated valance
charges within GGA+\emph{U} are listed in Table \ref{bader} together
with the LCAO (Ref. \cite{Evarestov}) and PW91 (Ref.
\cite{AKotomin}) results for comparison. Our present Bader analysis
gives the valency of U$^{1.71+}$N$^{1.71-}$, in qualitative
agreement with the LCAO (U$^{1.58+}$N$^{1.58-}$) and PW91
(U$^{1.66+}$N$^{1.66-}$) results. This agreement shows that the
ionicity in the U-N bond is intrinsic and therefore insensitive to
the different choices of computation methods. The dominant ionic
contribution to the U-N bond can also be seen from the total charge
density, which is plotted in Fig. \ref{charge}.

\subsection{Phonon dispersion curve of UN}
\begin{figure}[ptb]
\begin{center}
\includegraphics[width=0.8\linewidth]{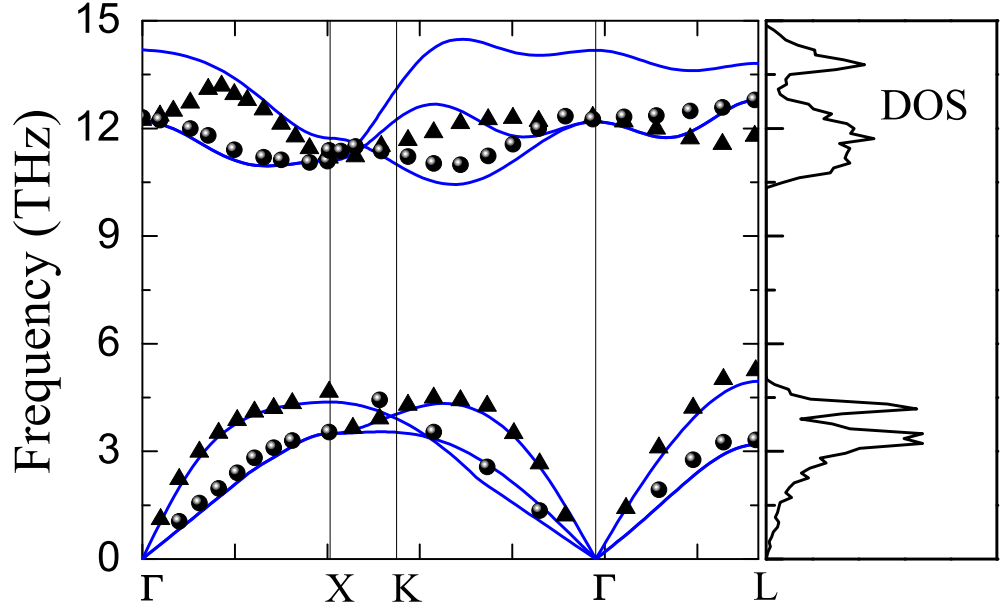}
\end{center}
\caption{(Color online) Calculated phonon dispersion curves (left
panel) and corresponding DOS (right panel) within GGA+\emph{U} approach at \emph{U}=2 eV for AFM UN. }%
\label{phonon-un}%
\end{figure}

Through the above discuss on atomic and electronic structures on
\emph{U}, we choose the GGA+\emph{U} approach with the Hubbard
\emph{U}= 2 eV to calculate the phonon dispersions for UN. In
calculating the phonon dispersion curves and the phonon density of
states, the Hellmann-Feynman theorem and the direct method
\cite{Parlinski} are employed. For the BZ integration, the
3$\times$3$\times$3 Monkhorst-Pack \emph{k}-point mesh is used for
the 2$\times$2$\times$2 UN supercell containing 64 atoms. In order
to calculate the Hellmann-Feynman forces, we displace two atoms (one
U and one N atoms) from their equilibrium positions and the
amplitude of all the displacements is 0.03 \AA. Besides, we have
calculated the Born effective charges of UN for their critical
importance in correcting the LO-TO splitting. Because of its high
symmetry for UN, the off-diagonal elements of the Born effective
charge tensor are all zero and the three diagonal elements
\emph{Z$_{xx}$}, \emph{Z$_{yy}$} and \emph{Z$_{zz}$} are the same.
Therefore, only \emph{Z$_{xx}$} is shown here. Our calculated
results for UN are $Z_{\mathrm{{U}}}^{\ast}$=+1.95 and
$Z_{\mathrm{{N}}}^{\ast}$=$-$1.95. The calculated phonon dispersion
curves along $\Gamma-X-K-\Gamma-L$ directions is displayed in Fig.
\ref{phonon-un}. The experimental data from Ref. \cite{Jackman} (at
T=4.2 K) are also presented for comparison. For NaCl type UN, there
are only two atoms in its formula unit, therefore, six phonon modes
exist in the dispersion relations. As shown in Fig. \ref{phonon-un},
our calculated LA/TA branch is in good agreement with experiment.
The remarkable splitting between LO and TO at $\Gamma$ point can be
attributed to the inclusion of the Born effective charges in our
phonon dispersion calculation. The TO frequency at $\Gamma$ point is
12.19 THz. This result is well consistent with the available
experimental value of 12.3 THz at 4.2 K. In addition, the phonon DOS
splits into two parts with one part in range of 0-4.7 THz where the
vibrations of uranium atoms are dominant and another part in the
domain of 10.5-15 THz where the vibrations mainly come from nitride
atoms. This evident gap between the optic modes and the acoustic
branches is because of the fact that the uranium atom is heavier
than nitride atom. In the following discussions, the reliability of
the phonon dispersion calculation will give an accuracy evaluation
of the thermodynamic properties.

\subsection{Thermodynamic properties}
To calculate thermodynamical quantities such as the lattice
vibration energy, thermal expansion, and specific heat, the
Helmholtz free energy \emph{F} in QHA is investigated as follows:
\begin{align}
F(V,T)=E(V)+F_{ph}(V,T)+F_{ele}(V,T),%
\end{align}
where \emph{E}(\emph{V}) stands for the ground state energy,
\emph{F$_{ph}$}(\emph{V,T}) is the phonon free energy at a given
unit cell volume \emph{V}, and \emph{F$_{ele}$} is electron
excitation energy. Under QHA, the \emph{F$_{ph}$}(\emph{V,T}) can be
calculated from phonon DOS by
\begin{align}
F_{\mathrm{ph}}(V,T)=k_{\mathrm{B}}T \int^{\infty}_{0}g(\omega)\ln[2sinh(\frac{\hbar\omega}{2k_{B}T})]d\omega,%
\label{pho}
\end{align}
where $\omega=\omega(V)$ denotes the volume-dependent phonon
frequencies, \emph{g}($\omega$) is the phonon density of states,
$\hbar$ is the Planck constant, and \emph{k$_{B}$} is the Boltzmann
constant. Equation (\ref{pho}) contains some effect of unharmonics
since the phonon frequencies have to be derived each time at the
current crystal volume \emph{V}. In addition, the specific heat for
constant volume \emph{C}$_{V}$ can be obtained directly as
\begin{align}
C_{V}&=(\frac{\partial F}{\partial
T})_{V}=k_{\mathrm{B}}\int_{\mathrm{0}}^{\mathrm{\infty}}d\omega
g(\omega)(\frac{\hbar \omega}{k_{\mathrm{B}}T})^{2} \frac{exp(\frac{\hbar \omega}{k_{\mathrm{B}}T})}{[exp(\frac{\hbar \omega}{k_{\mathrm{B}}T})-1]^{2}}.%
\end{align}
Then the specific heat at constant pressure \emph{C$_{P}$} is given
by
\begin{align}
C_{P}-C_{V}&=\alpha_{V}^{\mathrm{2}}(T)B(T)V(T)T,%
\end{align}
where the constant volume thermal expansion $\alpha_{V}$ is defined
by $\alpha_{V}=\frac{1}{V}(\frac{\partial V}{\partial T})_{P}$. The
electronic excitation effect on the specific heat is accounted by
the free-electron Fermi gas model, \emph{C$_{e}$}=$\gamma$$T_{e}$,
where $\gamma$ is the electronic specific heat coefficient. For
noninteracting electrons, the value of $\gamma$ is reasonable at low
electron temperature ($T_{e}$$<$3000 K). It is proportional to the
total density of states \emph{N(E$_{F}$)} at the Fermi level and is
given by $\frac{\pi^{2}}{3}k_{B}^{2}N(E_{F})$.

\begin{figure}[ptb]
\begin{center}
\includegraphics[width=0.5\linewidth]{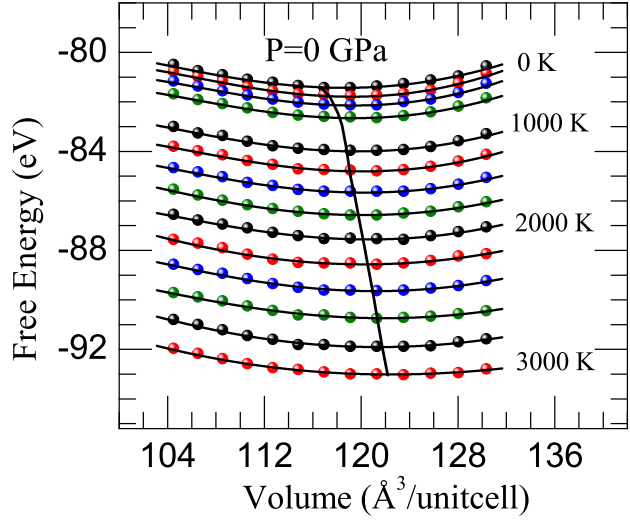}
\end{center}
\caption{(Color online) Dependence of the Helmholtz free energy
\emph{F}(\emph{T,V}) on crystal volume at various
temperatures. The locus of the minimum of the free energy for UN is also presented.}%
\label{helmholtz}%
\end{figure}

\begin{figure}[ptb]
\begin{center}
\includegraphics[width=0.6\linewidth]{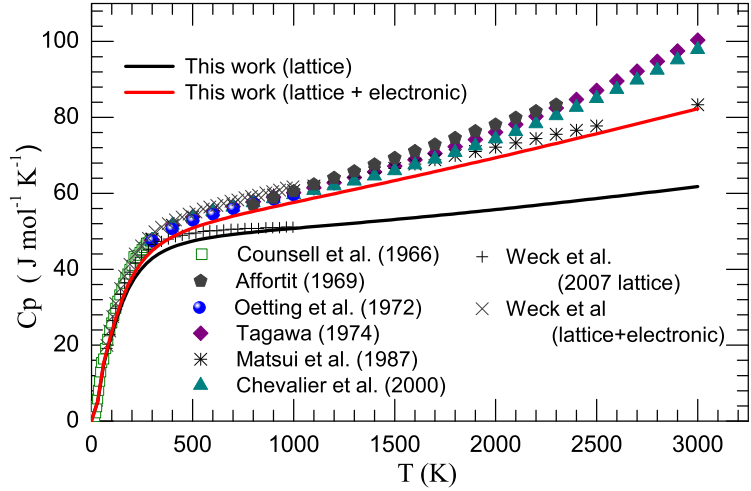}
\end{center}
\caption{(Color online) Heat capacity of UN calculated within
GGA+\emph{U} at \emph{U}=2 eV with and without considering the
contribution of electrons. Experimental data from Refs
\cite{Counsell,Affortit1,Affortit2,Oetting,Tagawa,Matsui} and
theoretical results from Refs. \cite{Chevalier,Philippe} are also displayed for comparison.}%
\label{cp}%
\end{figure}

In our calculations, the unit cells are expanded and compressed to a
set of constant volumes in calculations of free energy. Then the
equilibrium volume at temperature \emph{T} is obtained by minimizing
the free energy. The calculated free energy versus volume curves for
a number of selected temperatures is plotted in Fig.
\ref{helmholtz}, from which the volume expansion upon the
temperature increase can be derived. The calculated heat capacity
C$_{P}$ of UN is displayed in Fig. \ref{cp}. For comparison, the
experimental data from Refs.
\cite{Counsell,Affortit1,Affortit2,Oetting,Tagawa,Matsui} and the
theoretical results by Chevalier \emph{et al}. (N-U modelling
calculation) \cite{Chevalier} and Weck \emph{et al.} \cite{Philippe}
(all-electron calculation) are also presented. As shown in Fig.
\ref{cp}, the calculated thermodynamic functions with no electronic
excitation contribution at low temperature are in good agreement
with the experimental data up to around 200 K. However, in the high
temperature domain (\emph{T}$>$ 200 K), our calculated results with
only lattice vibration included evidently underestimate the C$_{P}$
compared with experimental values. This kind of underestimation in
the high temperature domain has also been observed in the
all-electron calculation \cite{Philippe}. Therefore, one needs to
take into account the conduction electrons contributions to the
\emph{C$_P$} for metallic material UN. Our estimated value for the
electronic specific heat coefficient $\gamma$ is equal to 26.7 mJ
K$^{-2}$ mol$^{-1}$. Although this value is somewhat lower than the
experimental value of 49.6 mJ K$^{-2}$ mol$^{-1}$ \cite{Holden}, it
is apparently larger than the value of 3.1 mJ K $^{-2}$ mol$^{-1}$
for its isostructural analog ThN \cite{Novion}. This can be
contributed to the unfilled 5\emph{f} electrons for UN. The specific
heat capacity including electronic contribution is also displayed in
Fig. \ref{cp}. One can see that the \emph{C$_P$} with electronic
corrections is largely enhanced, in agreement with the experimental
values within a broad temperature domain. Therefore, the partially
itinerant U 5\emph{f} electrons not only play a great role in
electronic properties, but also have considerable influence on the
thermodynamic performances of the intermetallic compounds such as
the present UN.

\section{conclusion}
In summary, we perform systematic first-principles calculations on
the structural, electronic, and thermal properties of UN using the
LDA/GGA+\emph{U} method. With the Hubbard \emph{U} correction, the
antiferromagnetic nature of UN is successfully predicted. The atomic
structure, including lattice parameters and bulk modulus can be
reasonably given, compared with corresponding experimental values.
The calculated electronic density of states shows the important role
that the 5\emph{f} electrons play in the conduction band as well as
in the valence band. By choosing \emph{U}=2 eV within GGA, the
phonon dispersions and phonon density of states can be reasonably
derived with regard to the experimental data. Using the reliable
phonon spectrum, the specific heat for constant pressure
\emph{C}$_{P}$ of UN including both lattice and conduction electron
contributions are calculated, the results of which are in good
agreement with experimental data. We expect that our calculated
results will be useful for the application of uranium nitride in the
Generation-IV reactor and nuclear industry.

\begin{acknowledgments}
This work was supported by NSFC under Grants No. 90921003 and No.
60776063.
\end{acknowledgments}


\begin{thebibliography}{99}
\bibitem{Proc}Proc. Global  Future Reactor Technologies (Tsukuba, Japan, Oct. 2005)

\bibitem {Matzke1992}H.J. Matzke, Diffusion Processes in Nuclear
Materials, (1992, Amsterdam: Elsevier)

\bibitem{Evarestov}R.A. Evarestov, A.I. Panin, A.V. Bandura, M.V. Losev, J. Phys, \textbf{117} (2008)
012015.

\bibitem{Evarestov2}R.A. Evarestov, M.V. Losev, A.I. Panin, N.S. Mosyagin, A.V. Titov,
Phys. Stat. Sol. (b) \textbf{245}, No. 1, 114¨C122 (2008)

\bibitem{Rafaja}D. Rafaja, L. Havela, R. Ku.el, F. Wastin, E. Colineau, T.Gouder,
Journal of Alloys and Compounds \textbf{386} (2005) 87¨C95.

\bibitem{Kotomin} E.A. Kotomin, R.W. Grimes, Y. Mastrikov, N.J. Ashley,
J. Phys.: Condens. Matter \textbf{19} (2007) 106208.

\bibitem{Doorn}C.F. van Doorn, P.de V. DuPlessis, J. Magn.
Magn. Mater. \textbf{5} (1977) 164.

\bibitem{antiferro}R. Tro\'{c}, J. Solid State Chem. \textbf{13} (1975) 14.

\bibitem{Dudarev}S.L. Dudarev, G.A. Botton, S.Y. Savrasov, C.J. Humphreys, A.P. Sutton, Phys. Rev. B \textbf{57} (1998) 1505.

\bibitem{Sun}B.Sun, P. Zhang, X.-G. Zhao, J. Chem. Phys. \textbf{128} (2008) 084705.

\bibitem{Wangbt}B.-T. Wang, H. Shi, W.-D. Li, P.
Zhang, Phys. Rev. B, \textbf{81} (2010) 045119.

\bibitem{Shi}H. Shi, P. Zhang, S.-S. Li, B. Sun, B. Wang,
Phys. Lett. A, \textbf{373} (2009) 3577.

\bibitem{G.Kresse1}G. Kresse, J. Furthm¨¹ller, computer code VASP, Vienna, (2005).

\bibitem{G.Kresse2}G. Kresse, J. Furthm¨¹ller, Phys. Rev. B \textbf{54} (1996) 11169.

\bibitem {PAW}P. E. Bl\"{o}chl, Phys. Rev. B \textbf{50} (1994) 17953.

\bibitem{LDA}W. Kohn, L.J. Sham, Phys. Rev. \textbf{140} (1965)
A1133.

\bibitem {GGA}J.P. Perdew, K. Burke, Y. Wang, Phys. Rev. B
\textbf{54} (1996) 16533.

\bibitem{Monkhorst}H.J. Monkhorst, J.D. Pack, Phys. Rev. B \textbf{13} (1976) 5188.

\bibitem{Brich}F. Brich, Phys. Rev. \textbf{71} (1947) 809.

\bibitem{Cordfunke}E.H.P. Cordfunke, J. Nucl. Mater.\textbf{56} (1975) 319.

\bibitem{Matzke}H.J. Matzke, Science of Advanced LMFBR Fuels (Amsterdam:
North-Holland), 1986.

\bibitem{Olsen}J.S. Olsen, L. Gerward, U. Benedict, J. Appl.
Crystallogr. \textbf{18} (1985) 37.

\bibitem{Petit}L. Petit, A. Svane, Z. Szotek, W.M. Temmerman, G.M.
Stocks, Phys. Rev. B, \textbf{80} (2009) 045124.

\bibitem{Norton}P.R. Norton, R.L. Tapping, D.K. Creber, W.J.L. Buyers, Phys. Rev. B. \textbf{21} (1980) 6.

\bibitem{Bader}R. Bader, Atoms in Molecules: A Quantum Theory, Oxford University Press, New York, (1990).

\bibitem{AKotomin}E.A. Kotomin, R.W. Grimes, Y. Mastrikov, N.J. Ashley, J. Phys.: Condens. Matter \textbf{19} (2007)
106208.

\bibitem{Parlinski}K. Parlinski, Z.Q. Li, Y. Kawazone, Phys. Rev. Lett, \textbf{78} (1997) 4063.

\bibitem{Jackman}J.A. Jackman, T.M. Holden, W.J.L. Buyers,
P.de V. DuPlessis, O. Vogt, J. Genossar, Phys. Rev. B \textbf{33}
(1986) 10.

\bibitem{Counsell}J.F. Counsell, R.M. Dell, J.F. Martin, Trans.
Faraday Soc. \textbf{62} (1966)

\bibitem{Affortit1}C. Affortit, High Temp - High Press. \textbf{1} (1969) 27.

\bibitem{Affortit2}C. Affortit, J. Nucl. Mater. \textbf{34} (1970) 105.

\bibitem{Oetting}F. Oetting, J.M. Leitnaker, J. Chem. Thermodyn. \textbf{4} (1972) 199.

\bibitem{Tagawa}H. Tagawa, J. Nucl. Mater. \textbf{ 51} (1974) 78.

\bibitem{Matsui}T. Matsui, R.W. Ohse, High Temp - High Press. \textbf{19} (1987) 1.

\bibitem{Chevalier}P.-Y. Chevalier, E. Fischer, B. Cheynet, J. Nucl. Mater. \textbf{280} (2000) 136-150.

\bibitem{Philippe}P.F. Weck, E. Kim, N. Balakrishnan, F. Poineau, C.B. Yeamans, K.R. Czerwinski, Chem.
Phys. Lett., \textbf{443} (2007) 82-86.

\bibitem{Holden}T.M. Holden, W.J.L. Buyers, C. Svensson, Phys.
Rev. B \textbf{30} (1984) 114.

\bibitem{Novion}C.de Novion, G. Costa, CR Acad. Sci. Ser. B \textbf{270}
(1970) 1415.

\end{thebibliography}
\end{document}